\documentclass{osa-article}

\usepackage[figuresright]{rotating}
\usepackage{graphicx}
\usepackage{amsmath}
\usepackage{enumerate}
\usepackage{cite}
\usepackage{subfigure}
\usepackage{booktabs}
\usepackage[T1]{fontenc}
\usepackage[misc]{ifsym} 
\usepackage{pifont}

\newenvironment{backmatter}{%
  \small%
  \newcommand{\bmsection}[1]{\par\medskip\noindent{\bfseries ##1.\enspace}}%
}{}

\journal{oe}

\articletype{Research Article}

\begin{document}

\title{A Novel Approach to Reducing Information Leakage for Quantum Key Distribution}

\author{Hao-Kun Mao,\authormark{1,6} Qiang Zhao,\authormark{2,6} Yu-Cheng Qiao,\authormark{3} Bing-Ze Yan,\authormark{1} Bing-Jie Xu,\authormark{4} Ahmed A. Abd EL-Latif, \authormark{5,1} Qiong Li,\authormark{1,*}}

\address{\authormark{1} Department of Computer Science and Technology, Harbin Institute of Technology, Harbin 150080, China \\
\authormark{2} College of software engineering, ZhengZhou University of Light Industry, ZhengZhou 450053, China \\
\authormark{3} Guangxi Key Lab Cryptography \& Information Security, Guilin University of Electronic Technology, Guilin 541004, Guangxi, China \\
\authormark{4} Science and Technology on Security Communication Laboratory, Institute of Southwestern Communication, Chengdu, 610041, China \\
\authormark{5} Department of Mathematics and Computer Science, Faculty of Science, Menoufia University, Shebin El-Koom, Egypt \\
\authormark{6} These authors contribute equally to this work \\
}

\email{\authormark{*}qiongli@hit.edu.cn} 



\begin{abstract}
Quantum key distribution (QKD) is an important branch of quantum information science as it holds promise for unconditionally secure communication. For QKD research, a central issue is to improve the final secure key rate (SKR) and the maximal transmission distance. To address this issue, most works focused on reducing the information leakage of QKD. In this paper, we propose a novel approach to further reduce the information leakage by specially considering the overlap between the information leakage of quantum part and post-processing part. The overlap means that the information leakage of post-processing part caused solely by multi-photon pulses is considered twice in previous studies, i.e., both in quantum part and post-processing part. Since the information carried by multi-photon pulses has been considered as completely known by Eve through the photon-number-splitting attack in quantum part, there is no need to consider it in post-processing part repetitively during the SKR calculation. Therefore, our approach can theoretically reduce the information leakage of a QKD protocol. Based on this idea, we derive the formulas to calculate the amount of information leakage for decoy-BB84 and sending-or-not-sending twin-field protocols. Simulation results for these two typical protocols also demonstrate that our approach evidently improves the SKR as well as the maximal transmission distance under practical experimental parameters.
\end{abstract}

\section{Introduction}
\label{sec:1}
Quantum key distribution (QKD) constitutes a promising solution for distributing unconditionally secure keys between two remote parties, such as Alice and Bob, in the presence of an eavesdropper, usually called Eve. Since the first QKD protocol, commonly known as the BB84 protocol, was proposed in 1984\cite{1_bennett1984quantum}, security analysis has been the focus of QKD studies. Though the ideal BB84 protocol has been proven to be unconditionally secure\cite{2_shor2000simple}, imperfect practical devices might still introduce security vulnerabilities, threatening the security of a practical QKD system. For instance, through the photon-number-splitting (PNS) attack \cite{3_lutkenhaus2000security, 4_brassard2000security, 5_lutkenhaus2002quantum} against the imperfect photon sources, Eve was capable of obtaining the complete information of each multi-photon pulse without causing any change in the quantum bit error rate (QBER). To address this challenge, Gottesman-Lo-Lutkenhaus-Preskill (GLLP) \cite{6_gottesman2004security} proved the unconditional security of a practical QKD system with imperfect devices. In the GLLP theory, each pulse can be represented as a mixed state of Fock states after phase randomization and the pulses can be classified based on the photon number. The information of all multi-photon pulses is assumed to be completely obtained by Eve, and the secure keys are solely generated from vacuum and single-photon pulses. Based on the GLLP theory, the secure key rate (SKR) of some discrete-variable (DV) QKD protocol, such as decoy-BB84 \cite{7_lo2005decoy, 8_wang2005beating, 9_wang2005decoy} and sending-or-not-sending twin-field (SNS-TF)\cite{10_lucamarini2018overcoming, 11_wang2018twin,12_wang2019beating}, can be calculated. In this paper, we point out that there is still possible to improve the SKR via estimating the information leakage during the information reconciliation (IR) more accurately.

Assume that the random variables $A$ and $B$ represent the sequences of Alice and Bob to be reconciled of length $N$, respectively. According to the noiseless coding theorem \cite{20_slepian1973noiseless}, the lower bound of the exchanged information $L_{all}$ for reliable IR can be calculated by the conditional entropy $H(A|B)$. In a DV-QKD system, $H(A|B)$ can be written as $Nh(e)$ \cite{21_martinez2015demystifying}, where $h(e) = - e{\log_2}(e) - (1 - e){\log_2}(1 - e)$ and $e$ is referred to as QBER. For a practical implementation of IR, $L_{all}$ is usually higher than $Nh(e)$. To this end, the IR efficiency $f = {L_{all}}/Nh(e)$ is introduced, while a smaller $f$ implies a better IR, and $f = 1$ represents the perfect IR \cite{21_martinez2015demystifying,31_tomamichel2017fundamental}. The commonly used IR protocols in DV-QKD systems can be generally divided into two categories \cite{22_mao2019high}: interactive and non-interactive. As the most widely used interactive IR protocol, Cascade \cite{21_martinez2015demystifying} detects and corrects errors by comparing the parity bits and performing binary search operations, respectively. The IR efficiency of Cascade is capable of approaching 1.02 \cite{23_pacher2015information}, but its high communication overhead potentially limits its practical application in QKD systems. Accordingly, the non-interactive IR protocols, based on low-density-parity-check (LDPC) \cite{22_mao2019high, 24_kiktenko2017symmetric} or polar codes \cite{25_yan2018improved, 26_kiktenko2020blind}, were proposed. For practical implementations of non-interactive protocols, the efficiencies typically range from 1.1 to 1.2 \cite{22_mao2019high}. In previous literatures about IR, all the exchanged information $L_{all}$ was considered as leaked information and subtracted directly from candidate secure keys during the SKR calculation.

However, we find out that only one part of $L_{all}$ needs to be subtracted during the SKR calculation. We notice that all the information of multi-photon pulses is assumed to be completely known by Eve through the PNS attack before IR. Thus, the exchanged information of IR caused solely by multi-photon pulses $L_M$ has also been known by Eve before IR and unable to provide any extra information for Eve after IR. Since the information of multi-photon pulses has been subtracted when analyzing the quantum part during the SKR calculation, $L_M$ is unnecessary to be subtracted again when analyzing the information leakage of IR of post-processing part. Following this idea, the SKR can be improved theoretically via avoiding the repetitive subtraction of $L_{M}$ during the SKR calculation. 

The rest of this paper is organized as follows. Sec. \ref{sec:2} presents the main idea of our approach. In Sec. \ref{sec:3}, the information leakage of two typical QKD protocols are analyzed. The performances of our approach are reported and analyzed through numerical simulations in Sec. \ref{sec:4}. Some conclusions are drawn in the last Section.

\section{The main idea of our approach}
\label{sec:2}
\begin{figure}[ht]
	\centering
	\includegraphics[width=\textwidth]{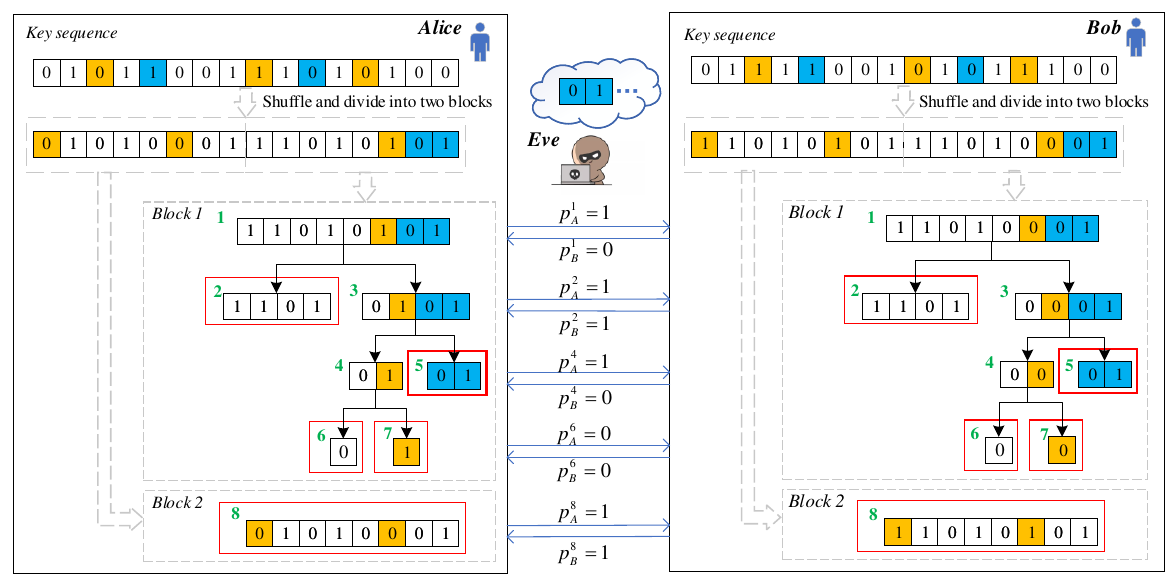}
	\caption{\textbf{A schematic workflow of Cascade protocol.} The keys that are inconsistent in two parties and generated from multi-photon pulses are filled with \textbf{orange} and \textbf{blue}, respectively. All the blocks (sub-blocks) generated during the error correction process are numbered in green, and $p_A^i$, $p_B^i$ represent the parity (i.e., the sum modulo 2 of all bits) of the \textit{ith} block belonging to Alice and Bob, respectively. }
	\label{fig:1}
\end{figure}

In this section, we take the Cascade IR protocol as an example to further elaborate on the main idea of our approach. As illustrated in Fig. \ref{fig:1}, the key sequences of Alice and Bob are first randomly shuffled and divided into two blocks of length $8$. Then, the parities of “Block 1” and “Block 2” of both parties are exchanged and compared simultaneously.

For "Block 1" with $p_A^1 \ne p_B^1$, a binary searching (i.e., recursively splitting a bit sequence into two halves and check the parities of the first halves of both parties) is performed to correct one error. Note that there is no need to exchange the parity of the second half since it can be easily deduced. For example, $p_A^3$ can be calculated by $p_A^1 \oplus p_A^2$. After four rounds of communications, the error is eventually found in sub-block 7. Through the exchanged parities, Eve can gain some information about the key sequences. For instance, the number of possible combinations of Alice’s sub-sequence in sub-block 2 decreases from 16 to 8 with the help of $p_A^2=1$. Overall, Eve obtains 1-bit information from each sub-block boxed in red and the total exchanged information of "Block 1" is 4-bit. However,  when considering the overlap between the information leakage of quantum part and post-processing part, the  effective amount of information $L_v$ obtained by Eve from IR is not equal to 4-bit. Let us suppose that there is a special sub-block 5 consisting two bits generated from multi-photon pulses. Before IR, Eve can obtain all the information of these two bits through the PNS attack. Though the parity of this sub-block is leaked to Eve during IR, Eve cannot acquire any extra useful information from the parity. In other words, the parity information has been known by Eve even if the parity is not exchanged. As a result, only the parity information obtained from the sub-block 2, 6, 7 are useful for Eve. Thus, $L_v$ equals to 3-bit, even though 4-bit information has been exchanged during IR.  

Since $p_A^8=p_B^8$, we do not deal with “Block 2” at this moment, even if this block still contains an even number of errors which cannot be detected or corrected until the subsequent processing. After several iterative passes of error correction (i.e. restarting shuffling and binary searching in each pass), the probability of correcting all errors is rather high, signifying a successful error correction. In addition, Cascade involves backtracking operations to benefit the binary searching, thereby reducing the information leakage. 

The above analysis suggests that the information leakage of a sub-block which solely contains bits generated from multi-photon pulses is not effective for Eve. Furthermore, the amount of effective information leakage $L_v$ to Eve can be obtained by estimating the number of such blocks, i.e., $L_v = L_{all} - L_M$.

\section{Information leakage analysis for two typical QKD protocols}
\label{sec:3}
Owing to the symmetry of DV-QKD protocols, without loss of generality, we only focus on the Alice's sequence $A$. Let $A_0$, $A_1$, $A_M$ represent the set of bits generated from the vacuum, single-photon and multi-photon pulses, respectively, s.t. $A = {A_0} \cup {A_1} \cup {A_M}$. No matter which IR protocol is applied, a collection of blocks $C= \{ c|c \subset A\}$ will be generated after IR, and each block c($c \in C$) is also a set containing several bits from $A$. Considering that each block $c$ is regarded to leak 1-bit information and some blocks from $C$ may be linearly correlated \cite{23_pacher2015information}, the original information leakage $L_{all} \le \left| C \right|$. Let ${C_{M}} = \{ c|(\forall c \in {C}) \wedge (c \subset {A_M})\}$, we have ${L_v} = {L_{all}} - {L_{M}} \le \left| {C} \right| - \left| {{C_M}} \right|$. 

Let $D$ represent the set of block lengths after IR (e.g. $D = \left\{ {8,4,2,1} \right\}$ in the case shown in Fig. \ref{fig:1}) and the sets $C^l, C_{M}^l (l \in D)$ satisfy ${C^l} = \{ c|(\forall c \in C) \wedge (\left| c \right| = l)\}$, $ C_{M}^l = \{ c|(\forall c \in {C_{M}}) \wedge (\left| c \right| = l)\}$. Known that the sequences are shuffled before each pass and the bits are then uniformly distributed in the sequences, we have
\begin{equation}
	\label{eq:1}
	\begin{split}
	{L_{v}}    &\le \left| {C} \right| - \left| {{C_M}} \right| \\
                         &= \sum\limits_{l \in D} {\left| {{C^l}} \right|}  - \sum\limits_{l \in D} {\left| {C_M^l} \right|} \\
		            &= \sum\limits_{l \in D} {\left| {{C^l}} \right|\left[ {1 - {{\left({\Delta _{M}} \right)}^l}} \right]} \\
			      &\le \sum\limits_{l \in D} {\left| {{C^l}} \right|\left[ {\min \left( {1 - {{\left(\Delta _{M}^{\inf } \right)}^l},1} \right)} \right]}, \\
	\end{split}
\end{equation}
where $\Delta _{M}$ represents the proportion of the count rate of the multi-photon pulses to that of the signal pulses. 

Let $R$ represents the SKR per pulse in a practical DV-QKD system and $q$ is the sifting efficiency, we eventually get the formula for calculating the improved $R$ as
\begin{equation}
	\label{eq:2}
	R = q{Q_\mu }\max \left\{ {\Delta _1^{\inf }\left[ {1 - h(e_1^{\sup })} \right] + \Delta _0^{\inf } - \sum\limits_{l \in D} {\left( {\left| {{C^l}} \right|/N} \right)\left[ {\min \left( {1 - {{\left( \Delta _{M}^{\inf } \right)}^l},1} \right)} \right]} ,0} \right\}.
\end{equation}

We can see from Eq. (\ref{eq:1}) that the critical parameter is $\Delta _{M}^{\inf }$. For this reason, we deduce the methods of calculating $\Delta _{M}^{\inf}$ for two typical DV-QKD protocols based on the GLLP theory, that is, decoy-BB84 and SNS-TF protocols. Note that for some modified TF-type protocols, such as no-phase-postselection protocol \cite{27_cui2019twin}, phase-matching protocol \cite{28_ma2018phase}, etc., our new approach cannot be directly applied, as the signals can not be directly classified based on the number of photons in these protocols. For these protocols, additional analysis and proofs based on our main idea may be required.

\subsection{For decoy-BB84 protocol}
Known that  $\Delta _{M}^{\inf } = 1 - \Delta _0^{\sup } - \Delta _1^{\sup }$ in a decoy-BB84 protocol, we just need to focus on the calculation of $\Delta _0^{\sup }$ and $\Delta _1^{\sup }$. By using the decoy-state method, we have
\begin{equation}
	\label{eq:15}
	{Q_k} = \sum\limits_{i = 0}^\infty  {{Y_i}} \frac{{k^i}}{{i!}}{e^{ -k}}, k\in\{\mu,\nu_1,\nu_2\} \\
\end{equation}
where $Q_i$ represents the count rate of the pulse whose photon number is $i (i = 0,1,2,...,n)$ and the ratio ${\Delta _i} = {Q_i}/{Q_\mu }$.

Based on Eq. (\ref{eq:15}), we have
\begin{equation}
	\label{eq:16}
	{Q_{{\nu_1}}}{e^{{\nu_1}}} - {Q_{{\nu_2}}}{e^{{\nu_2}}} = {Y_1}({\nu_1} - {\nu_2}) + \sum\limits_{i=2}^\infty  {\frac{{{Y_i}}}{{i!}}(\nu_1^i}  - \nu_2^i) \ge {Y_1}({\nu_1} - {\nu_2}).
\end{equation}

Then the upper bounds of $Y_1$ and ${\Delta _1}$ can be obtained as follow:
\begin{equation}
	\label{eq:17}
	\begin{split}
	{Y_1} &\le \frac{{{Q_{{v_1}}}{e^{{\nu _1}}} - {Q_{{\nu_2}}}{e^{{\nu _2}}}}}{{{\nu _1} - {\nu _2}}}, \\
	{\Delta _1} &= \frac{{{Q_1}}}{{{Q_\mu }}} = \frac{{{Y_1}\mu {e^{ - \mu }}}}{{{Q_\mu }}} \le \frac{{({Q_{{\nu_1}}}{e^{{\nu _1}}} - {Q_{{\nu_2}}}{e^{{\nu _2}}})\mu{e^{ -\mu}}}}{{({\nu_1} - {\nu_2}){Q_\mu}}}. \\
	\end{split}
\end{equation}

For $Y_0$ and ${\Delta _0}$, according to Eq. (\ref{eq:15}), we have
\begin{equation}
	\label{eq:18}
	{Q_{{\nu_2}}}{e^{{\nu_2}}} = {Y_0} + {Y_1}{\nu_2} + \sum\limits_{i = 2}^\infty  {{Y_i}} \frac{{\nu_2^i}}{{i!}} \ge {Y_0} + {Y_1}{\nu_2},
\end{equation}

thus the upper bounds of $Y_0$ and ${\Delta _0}$ can be obtained as
\begin{equation}
	\label{eq:19}
	\begin{split}
	{Y_0} &\le {Q_{{\nu_2}}}{e^{{\nu_2}}} - {Y_1}{\nu_2} \le {Q_{{\nu_2}}}{e^{{\nu_2}}} - Y_1^{\inf }{\nu_2}, \\
	{\Delta _0} &= \frac{{{Q_0}}}{{{Q_\mu}}} = \frac{{{Y_0}{e^{ -\mu}}}}{{{Q_\mu}}} \le \frac{{({Q_{{\nu_2}}}{e^{{\nu_2}}} - Y_1^{\inf }{\nu_2}){e^{ -\mu}}}}{{{Q_\mu}}}. \\
	\end{split}
\end{equation}

Based on Ref. \cite{13_ma2005practical}, the estimated formula of $Y_1^{\inf }$ and $Y_0^{\inf}$ are given as follows. 
\begin{equation}
	\label{eq:19-1}
	\begin{split}
	Y_1^{\inf } &= \frac{\mu }{{\mu {v_1} - \mu {v_2} - v_1^2 + v_2^2}}\left( {{Q_{{v_1}}}{e^{{v_1}}} - {Q_{{v_2}}}{e^{{v_2}}} - \frac{{v_1^2 - v_2^2}}{{{\mu ^2}}}\left( {{Q_\mu }{e^\mu } - Y_0^{\inf }} \right)} \right)\\
	Y_0^{\inf } &= \max \left\{ {\frac{{{v_1}{Q_{{v_2}}}{e^{{v_2}}} - {v_2}{Q_{{v_1}}}{e^{{v_1}}}}}{{{v_1} - {v_2}}},0} \right\}\\
	\end{split}
\end{equation}
According to Eq. (\ref{eq:17}), (\ref{eq:19}), and (\ref{eq:19-1}), we can calculate the value of $\Delta _{M}^{\inf }$, thus an optimized SKR can be generated for the decoy-BB84 protocol.

\subsection{For SNS-TF protocol}
The original formula for calculating SKR in an SNS-TF protocol was given in Ref. \cite{11_wang2018twin}: 
\begin{equation}
	\label{eq:20}
	R = P_A^ZP_B^Z\left\{ {\left[ {{\varepsilon _A}\left( {1 - {\varepsilon _B}} \right){\mu _A}{e^{ - {\mu _A}}} + {\varepsilon _B}\left( {1 - {\varepsilon _A}} \right){\mu _B}{e^{ - {\mu _B}}}} \right]s_1^z\left[ {1 - h\left( {e_1^{ph}} \right)} \right] - {s_z}{f}h\left( {{E_z}} \right)} \right\},
\end{equation}
where the meanings of $P_A^Z$, $P_B^Z$, $\varepsilon _A$, $\mu _A$, $\varepsilon _B$, $\mu _B$ are given in Table \ref{tab:2}, $s_1^z$, $e_1^{ph}$ refer to the count rate and phase-error-rate in Z window respectively, $s_z$, $E_z$ refer to the total count rate and bit-error-rate in Z windows respectively. By using our new approach, we convert Eq. (\ref{eq:20}) to 
\begin{equation}
	\label{eq:21}
	R = P_A^ZP_B^Z\left\{ {\left[ {{\varepsilon _A}\left( {1 - {\varepsilon _B}} \right){\mu _A}{e^{ - {\mu _A}}} + {\varepsilon _B}\left( {1 - {\varepsilon _A}} \right){\mu _B}{e^{ - {\mu _B}}}} \right]s_1^z\left[ {1 - h\left( {e_1^{ph}} \right)} \right] - L} \right\}.
\end{equation}

Note that only pulses with single-photon from Z windows can generate secure keys. Assuming ${\Delta _{M}}$ to be the count rate of multi-photon pulses in Z windows, we have
\begin{equation}
	\label{eq:22}
	{\Delta _{M}} = 1 - {P_1}s_1^Z - {P_0}s_0^Z,
\end{equation}
where $s_0^Z$ is the count rate of vacuum pulse in Z windows and $P_0$, $P_1$ are the probability of the photon number equaling 0 or 1, respectively. By analyzing the different conditions of choosing sending or not-sending for Alice and Bob, we have

\begin{equation}
	\label{eq:23}
	\begin{split}
	{P_0} &= {\varepsilon _A}\left( {1 - {\varepsilon _B}} \right){e^{ - {\mu _A}}} + {\varepsilon _B}\left( {1 - {\varepsilon _A}} \right){e^{ - {\mu _B}}} + {\varepsilon _A}{\varepsilon _B}{e^{ - \left( {{\mu _A} + {\mu _B}} \right)}} + \left( {1 - {\varepsilon _A}} \right)\left( {1 - {\varepsilon _B}} \right), \\
	{P_1} &= {\varepsilon _A}\left( {1 - {\varepsilon _B}} \right){\mu _A}{e^{ - {\mu _A}}} + {\varepsilon _B}\left( {1 - {\varepsilon _A}} \right){\mu _B}{e^{ - {\mu _B}}} + {\varepsilon _A}{\varepsilon _B}\left( {{\mu _A} + {\mu _B}} \right){e^{ - \left( {{\mu _A} + {\mu _B}} \right)}}. \\
	\end{split}
\end{equation}

We then analyze $s_0^Z$ and $s_1^Z$. The parameter $s_0^Z$ can be directly obtained from ${S_{00}}$, which is the count rate of vacuum pulses in X windows:
\begin{equation}
	\label{eq:24}
	s_0^Z = {S_{00}}.
\end{equation}

For $s_1^Z$, its lower bound has been given in Ref. \cite{11_wang2018twin}, which can give us a revelation to estimate its upper bound:
\begin{equation}
	\label{eq:25}
	s_1^Z = \frac{{{\mu _{A1}}}}{{{\mu _{A1}} + {\mu _{B1}}}}s_{10}^z + \frac{{{\mu _{B1}}}}{{{\mu _{A1}} + {\mu _{B1}}}}s_{01}^z,
\end{equation}
where:
\begin{equation}
	\label{eq:26}
	s_{10}^Z \le \frac{{{e^{{\mu _{A2}}}}{S_{20}} - {e^{{\mu _{A1}}}}{S_{10}}}}{{{\mu _{A2}} - {\mu _{A1}}}},{\rm{ }}s_{01}^Z \le \frac{{{e^{{\mu _{B2}}}}{S_{02}} - {e^{{\mu _{B1}}}}{S_{01}}}}{{{\mu _{B2}} - {\mu _{B1}}}},
\end{equation}
where the meanings of all the above parameters are given in Ref. \cite{11_wang2018twin}. By substituting Eq. (\ref{eq:23}),(\ref{eq:24}), and (\ref{eq:25}) into Eq. (\ref{eq:22}), we finally get the formula of $\Delta _{M}^{\inf }$ for the SNS-TF protocol.

\section{Simulations and Discussions}
\label{sec:4}
We evaluate the performance improvement of our new approach over the original GLLP formula through simulations under practical experimental parameters. The key simulation parameters of the decoy-BB84 \cite{29_yuan201810} and SNS-TF protocols \cite{30_chen2020sending} are detailed in Table \ref{tab:1} and \ref{tab:2}, respectively. 

In addition, we can clearly see from Eq. (\ref{eq:1}) that the smaller $l$ is, the better our approach performs. However, in a commonly used non-interactive IR protocol, a larger $l$ is usually applied to achieve a higher $f$. Therefore, we consider that the non-interactive IR protocols cannot gain obvious SKR improvement from our approach. In contrast, a collection of blocks with different block lengths whose value can be as low as 1, will be obtained after Cascade. We thus conclude that a greater SKR improvement can be obtained when using Cascade. Therefore, we apply Cascade as the IR protocol in our simulations and its IR efficiency $f$ is set to the optimal value 1.
\begin{table}[ht]
\caption{Key simulation parameters for the decoy-BB84 protocol. $\mu$, $\nu_1$, $\nu_2$: average photon number for signal, decoy and vacuum pulses; $q$: sifting efficiency; $\alpha$: the channel loss; $d$: the dark count rate of the detector; $\eta_d$: detection efficiency of the detector; $e_{det}$: systematic errors.}
\label{tab:1}
\centering
\begin{tabular}{cccccccc}
\hline
$\mu$ & $\nu_1$ & $\nu_2$ & $q$ & $\alpha$ & $d$ & $\eta_d$ & $e_{det}$ \\
\hline
0.4 & 0.1 & 0.0007 & 0.9 & 0.20 dB/km & $10^{-5}$ & 20\% & 0.033 \\
\hline
\end{tabular}
\end{table}

\begin{table}[ht]
\caption{Key simulation parameters for the SNS-TF protocol. $P_A^Z$, $P_B^Z$: the probability of choosing the $Z$ window by Alice and Bob;  $\varepsilon _A$, $\varepsilon _B$, $\mu _A$,  $\mu _B$: the probability of choosing the Sending mode in $Z$ windows and the corresponding average photon number of each sending pulse by Alice and Bob; $e_d$: the misalignment error in the X window; The definitions of $\alpha$, $d$, $\eta_d$, $e_{det}$ are same as in Table \ref{tab:1}.}
\label{tab:2}
\centering
\begin{tabular}{ccccccccccc}
\hline
$P_A^Z$ & $P_B^Z$ & $\varepsilon _A$ & $\varepsilon _B$ & $\mu _A$ & $\mu _B$ & $e_d$ & $\alpha$ & $d$ & $\eta_d$ & $e_{det}$ \\
\hline
0.7 & 0.8 & 0.022 & 0.48 & 0.042 & 0.425 & 5\% & 0.20 dB/km & $10^{-10}$ & 50\% & 0.033 \\
\hline
\end{tabular}
\end{table}

\begin{figure}[ht]
	\centering
	\includegraphics[width=0.75\textwidth]{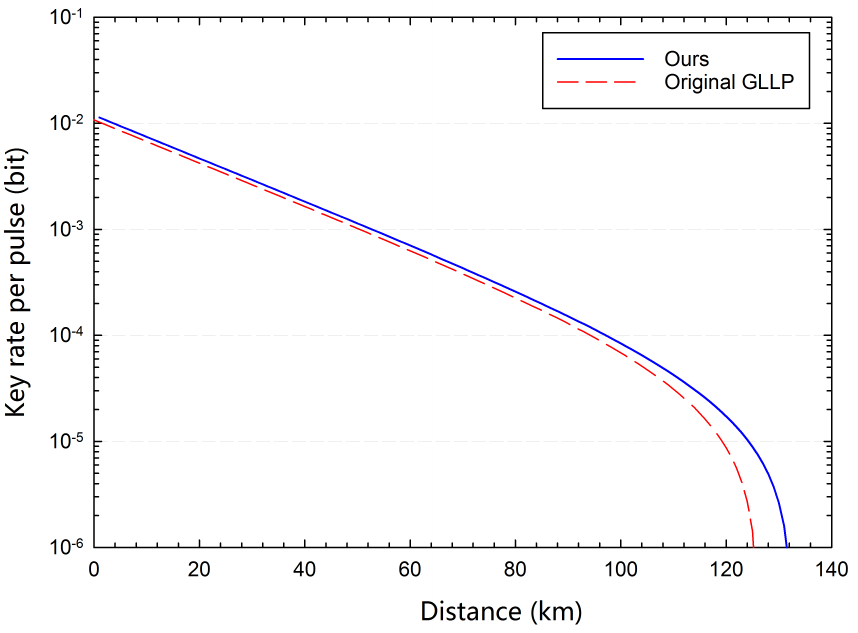}
	\caption{SKR per pulse vs. transmission distance with our improved and the original approach for a decoy-BB84 protocol \cite{29_yuan201810}.}
	\label{fig:2}
\end{figure}

\begin{figure}[ht]
	\centering
	\includegraphics[width=0.75\textwidth]{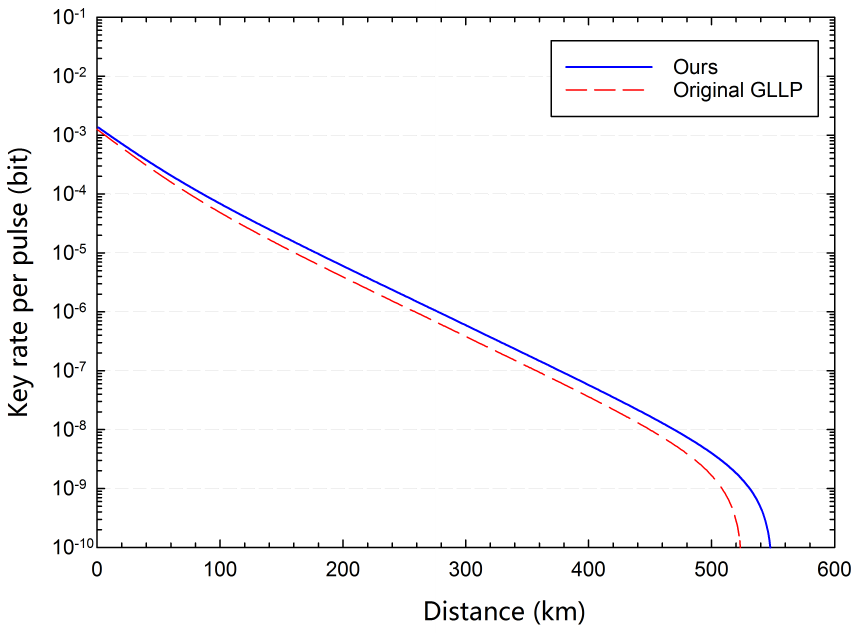}
	\caption{SKR per pulse vs. transmission distance with our improved and the original approach for a SNS-TF protocol \cite{11_wang2018twin}.}
	\label{fig:3}
\end{figure}

We show the numerical results for decoy-BB84 protocol and SNS-TF protocol in Fig. \ref{fig:2} and Fig. \ref{fig:3}. Simulation results show that our approach can not only improve the SKR at any distance, but also increase the maximal transmission distances of two protocols by 5km and 20km, respectively. We note that the SKR improvement of our approach does not depend on the performance boost of optical devices, but rather comes from the more accurate estimation of the information leakage of IR.

Though our approach can significantly improve the SKR at any distance, Eq. (\ref{eq:2}) needs further modification when our formula is expanded to the finite-size regime. Since the information leakage of our new approach is affected by finite-size effects, additional analysis may be needed. In addition, based on different protocols, related parameters can be optimized to obtain a better  $\Delta _{M}^{\inf }$, thus improving SKR.

\section{Conclusions}
In this study, we propose a novel approach to reduce the information leakage of IR by specially considering the overlap between the information leakage of quantum part and the post-processing part. Benefiting from avoiding the repetitive subtraction of information leakage caused by multi-photon pulses during IR, our approach theoretically improves the SKR of a QKD protocol. Simulation results for decoy-BB84 and SNS-TF protocols show that our approach is capable of improving both the SKRs at any distance and the maximal transmission distance. Note that our approach can be applied to all DV-QKD protocols based on the GLLP theory.

Besides, we consider that the main idea of our approach may have implications for continuous-variable (CV) QKD protocols as well. Our main idea implies that the quantum and classical signal processing are not completely independent. Thus, when estimating the leaked information through the classical channel, whether it overlaps with the information leaked in the quantum part also needs to be considered. The above analyses bring us an implication for CV-QKD system, that is, a better SKR can be calculated for a CV-QKD system if there exist similar situations in this system.

\section{Backmatter}
\begin{backmatter}
\bmsection{Funding}

\bmsection{Acknowledgments}
This work is supported by the National Natural Science Foundation of China (Grant Number: 62071151, 61301099). Special thanks goes to Dr. Xuan Wen for the helpful discussions.

\bmsection{Disclosures}
The authors declare no conflicts of interest.

\bmsection{Data Availability Statement}
Data underlying the results presented in this paper are not publicly available at this time but may be obtained from the authors upon reasonable request.

\end{backmatter}

\section{References}


\end{document}